\documentstyle[12pt,epsf]{article}

\input{psfig.sty}

\newcommand{\nc}{\newcommand}
\nc{\beq}{\begin{equation}}
\nc{\eeq}{\end{equation}}
\nc{\beqa}{\begin{eqnarray}}
\nc{\eeqa}{\end{eqnarray}}

\newwrite\ffile\global\newcount\figno \global\figno=1

\def\writedef#1{}

\def\figin{\epsfcheck\figin}\def\figins{\epsfcheck\figins}
\def\epsfcheck{\ifx\epsfbox\UnDeFiNeD
\message{(NO epsf.tex, FIGURES WILL BE IGNORED)}
\gdef\figin##1{\vskip2in}\gdef\figins##1{\hskip.5in}
\else\message{(FIGURES WILL BE INCLUDED)}%
\gdef\figin##1{##1}\gdef\figins##1{##1}\fi}

\def\figinsert{}
\def\ifig#1#2#3{\xdef#1{fig.~\the\figno}
\writedef{#1\leftbracket fig.\noexpand~\the\figno}%
\figinsert\figin{\centerline{#3}}\medskip\centerline{\vbox{\baselineskip12pt
\advance\hsize by -1truein\center\footnotesize{  Fig.~\the\figno.} #2}}
\bigskip\endinsert\global\advance\figno by1}
\def\endinsert{}

\begin{document}

\title{\large{\bf Flavor Gauge Bosons at the Tevatron}}

\author{
Gustavo Burdman$^{(a)}$\thanks{burdman@bu.edu},~~~ 
R. Sekhar Chivukula$^{(a)}$\thanks{sekhar@bu.edu}~~~and 
Nick Evans$^{(b)}$\thanks{n.evans@hep.phys.soton.ac.uk}
\\
\\
{\small\em (a) Department of Physics,
Boston University, Boston, MA 02215, USA.} \\ 
{\small\em (b) Department of Physics,
University of Southampton, Southampton, S017 1BJ, UK.} }

\date{ }

\maketitle

\begin{picture}(0,0)(0,0)
\put(350,260){BUHEP-00-8}
\put(350,245){SHEP-00-03}
\end{picture}

\begin{abstract}
  We investigate collider signals for gauged flavor symmetries that
  have been proposed in models of dynamical electroweak symmetry
  breaking and fermion mass generation.  We consider the limits on the
  masses of the gauge bosons in these models which can be extracted
  from Tevatron Run I data in dijet production.  Estimates of the Run
  II search potential are provided. We show that the models also give
  rise to significant signals in single top production which may be
  visible at Run II.  In particular we study chiral quark family
  symmetry and SU(9) chiral flavor symmetry. The Run I limits on the
  gauge bosons in these models lie between $(1.5-2)~$TeV and should
  increase to about $3$~TeV in Run II.  Finally, we show that an
  SU(12) enlargement of the SU(9) model, including leptonic
  interactions, is constrained by low energy atomic parity violation
  experiments to lie outside the reach of the Tevatron.
\end{abstract}

\newpage

\section{Introduction}

The origin of the Standard Model's (SM) familiar 
$SU(3)_c \times SU(2)_L \times U(1)_Y$ gauge symmetry remains theoretically
unclear. In the limit where we neglect all gauge
interactions and fermion masses, the fermion sector of the model 
possesses a large SU(45) global symmetry
corresponding to the fact that in this limit there are 45 chiral fermion
fields that are indistinguishable. The gauge interactions of the SM are
by necessity subgroups of this maximal symmetry but in principle a larger 
subgroup of this symmetry might be gauged and broken to the SM groups at 
high energies. 

Such gauged flavor symmetries have been invoked in a number of
scenarios to play a role in the dynamical generation of fermion
masses. For example they may play the part of extended technicolor
\cite{ETC} interactions in technicolor models \cite{TC} or top
condensation models \cite{tc}, feeding the electroweak symmetry (EWS)
breaking fermion condensate down to provide masses for the lighter
standard model fermions. Strongly interacting flavor gauge
interactions may also be responsible for the condensation of the
fermions directly involved in EWS breaking. For example, top
condensation has been postulated to result from a Topcolor gauge group
\cite{tcol} and in the model of \cite{king} from family gauge
interactions. There has been renewed interest in these models recently
with the realization that variants, in which the top mixes with
singlet quarks, can give rise to both the EW scale and an acceptable
top mass via a seesaw mass spectrum \cite{tseesaw}. These top seesaw
models have the added benefit of a decoupling limit which allows the
presence of the singlet fields to be suppressed in precision EW
measurements bringing these dynamical models in line with the data.
Flavor universal variants of the top-seesaw idea have been proposed 
in Ref.~\cite{fuseesaw},  where the dynamics is driven by
family or large flavor gauge symmetries.

The naive gauging of flavor symmetries at low scales (of order a few
TeV) often gives rise to unacceptably large flavor changing neutral
currents (FCNC) since gauge and mass eigenstates need not coincide.
For instance, gauge symmetries that give rise to direct contributions to
$K^0-\bar{K}^0$ mixing are typically constrained to lie above 500 TeV
in mass scale.  There are, however, models that survive these
constraints.  Gauge groups that only act on the third family are less
experimentally constrained - Topcolor \cite{tcol} 
is such an example. Models in
which the chiral flavor symmetries of the SM fermions are gauged
preserving the SM $U(3)^5$ \cite{ctsm} flavor symmetry can respect the
SM GIM mechanism and do not give rise to tree level FCNCs
\cite{georgi}.  In addition, there are also strong constraints on
gauged flavor models where the dynamics responsible for the breaking
of the flavor symmetry does not respect custodial isospin
\cite{mixing}.  We shall restrict ourselves to models where the top
mass is the sole source of custodial isospin breaking. In particular we
we will study a model where the SU(3) chiral family symmetry of the quarks
is gauged and another where the full SU(9) family-color multiplicity
of the quarks is gauged, corresponding to the models of \cite{fuseesaw}. 
In the spirit of these models it is also interesting to consider 
chiral flavor symmetries that include the leptons which might be expected 
to give interesting contributions to Drell-Yan production. The obvious
extension has a gauged SU(12) flavor symmetry but we show in the final section
that an analysis of low energy atomic parity violation experiments
places constraints on the gauge bosons of such models of order 10 TeV
and they are thus outside the reach of the Tevatron.

Since these new flavor interactions may exist at relatively low scales
(a few TeV) and may play an integral part in either EWS breaking or
fermion mass generation it is interesting to study current
experimental bounds on the corresponding gauge bosons.  In a previous
paper we investigated the limits from Z-pole precision measurements
\cite{zbounds}. Although the limits obtained vary across models, the
typical lower bound on the mass scale is $2$~TeV.  Here we study the
potential of direct searches at the Fermilab Tevatron collider. In
particular we study effects in dijet production (in the spirit
of the analysis in \cite{simmons, bertram}) and single top production.
When possible, we first
establish bounds from the existing Run~I data
(they are typically  1-2 TeV). We then project the sensitivity 
of the Tevatron in  Run~II and show the bounds are more than competitive
with the precision data bounds. If these gauge symmetries do have a 
role to play in EWS breaking then they must presumably be broken at scales 
close to the EW scale and these bounds therefore represent a significant
probe of the interesting parameter space. 

\section{Constraints on Models }

We present three  models of flavoron physics.  While this list is not
exhaustive, we believe these examples cover a broad range of signals
at the Tevatron collider. In what follows, only the couplings to
standard model fermions will be specified. Explicit models include
additional fermions, necessary for either flavor gauge symmetry
breaking and/or anomaly cancellation, which typically have masses of
order of the flavor gauge symmetry breaking scale.

\subsection{Chiral Quark Family Symmetry}

The gauging of the chiral family symmetry of the left handed quarks
has been motivated in technicolor \cite{georgi}, top condensate
\cite{king} and flavor universal see saw models \cite{fuseesaw}. The
minimal representative model has a gauged SU(3) family symmetry, in
addition to the SM interactions, acting on the three left handed
quark\footnote{One can also imagine the same symmetry acting on
  leptons~\cite{fuseesaw}.  Here we only consider the quarks since
  they lead to signals at hadron colliders.}  doublets $Q =((t,b)^i_L,
(c,s)^i_L, (u,d)^i_L )$ where $i$ is a QCD index which commutes with
the family symmetry~\cite{fuseesaw}.  We assume that some massive
sector completely breaks the SU(3) family gauge group to an global
SU(3) family symmetry, giving the family gauge bosons (``familons'')
masses of order $M_F = g_F V$ where $V$ is the mass scale associated
with the symmetry breaking. There is no mixing between the flavor and
standard model gauge bosons.  Note that with this gauge symmetry and
symmetry breaking pattern, the (approximate) SM $U(3)^5$ global
symmetry responsible for the GIM mechanism \cite{ctsm} remains and the
model is free of tree level FCNCs \cite{georgi}. The interactions of
the massive flavorons are summarized by the couplings
\begin{equation}
{\cal L} = i g_F A^{\mu a} \bar{Q} \gamma_\mu T^a  Q~,
\end{equation}
where $T^a$ are the generators of SU(3) symmetry acting on the three
families of left-handed quarks.

\noindent  
The SU(3) coupling $g_F$ cannot be too large or this interaction will
cause a chiral symmetry breaking condensate between the left-handed
ordinary fermions and right-handed fermions which must be present in
the theory to eliminate gauge anomalies. This would result in
TeV-scale fermion masses and a scale for electroweak symmetry breaking
which is too high. We may estimate the upper bound on $g_F$ by
approximating, at low energies, the interactions of the massive flavor
gauge bosons by a Nambu--Jona-Lasinio (NJL) model with the
four-fermion interaction
\begin{equation}
{\cal L}_{\rm eff} = - {2\pi\kappa_F \over M_{F}^2}  
\left( \sum_f \bar{Q} \gamma_\mu T^a Q \right)^2~,
\label{ucnjl}
\end{equation}
where $\kappa\equiv g_F^2/4\pi$.  Applying the usual NJL
analysis\footnote{Note that, defining the theory in terms of a
  momentum-space cutoff $\Lambda$, a four fermion interaction $G
  \bar{\psi}\psi \bar{\psi} \psi $ has a critical coupling $G_c = 2
  \pi^2/\Lambda^2$ \cite{NJL}.  }, we see that $\kappa_F$ cannot
exceed
\begin{equation}
\kappa_{crit} = {2 N \pi \over (N^2-1)} = 2.36~,
\label{kcsu3}
\end{equation}
where $N=3$ for chiral quark flavor symmetry.

\noindent
In Ref.\cite{zbounds} we obtained bounds on flavor gauge boson
masses from electroweak precision measurements. The lower bound
obtained for a critically coupled familon is $M_F > 1.9$~TeV, at
$95\%$ C.L.  Here we will investigate the reach of direct searches.
First, we consider the bounds from the existing Tevatron data.  As is
the case for the universal coloron model \cite{unicol}, stringent
limits will come from the study of the angular behavior of the dijet
cross section \cite{bertram}.  The contributions arising in the chiral
quark family model are the consequence of the exchange of the familon
gauge boson in the various possible channels. The resulting
modification of the quark scattering matrix elements are given in
Section A.2 of the Appendix.

\noindent
In Fig.\ref{f3vqcd} we plot the ratio of the dijet mass 
distribution for $|\eta|<0.5$ to the mass distribution  
with $0.5<|\eta|<1.0$, with $\eta$ the jet pseudo-rapidity. 
This ratio, as noted for instance in Refs.\cite{bertram,d0kdata}, 
is very sensitive to new physics producing effects concentrated in the 
central region, and in general affecting the angular distribution 
of dijets. Also it is expected that in this ratio there is a large cancellation
of uncertainties coming from softer QCD effects. 
The data points are from the D0 data in Ref.\cite{d0kdata}, and 
the error bars show the statistical and systematic
errors added in quadrature. 
The histogram corresponds to the QCD prediction, obtained to 
next-to-leading order (NLO) with the use of JETRAD (see \cite{bertram,d0kdata}
for details).
The familon contribution is known only at leading order (LO). Thus, in 
order to estimate their NLO dijet spectrum, we compute the fractional 
excess with respect to LO QCD and then multiply it by the NLO QCD 
result. 
We consider various familon masses, with the coupling set at its 
critical value. 

In order to obtain a lower mass limit we follow the 
procedure described in Ref.\cite{bertram}. 
We construct the Gaussian likelihood function
\begin{equation}
P(x)=\frac{1}{2\pi^2 det(S)} \exp{\left(-\frac{1}{2}[d-t(x)]^T S^{-1}
[d-t(x)]\right)}~,
\label{pdf}
\end{equation}
where the vector $d$ contains the data points in the various mass
bins, $t(x)$ is the vector of theoretical predictions for a given mass
and coupling $x=\kappa_F/M^2_F$; and $S$ is the covariant matrix.  To
obtain $95\%$ confidence level limits, we require
\begin{equation}
Q(x_{\rm max})\equiv \int_{0}^{x_{\rm max}}P(x) dx = 0.95 Q(\infty)~,
\label{cldef}
\end{equation} 
with $x_{\rm max}$ the value defining the mass bound. 
Making use of the the Run~I data 
we then obtain  mass bounds for the familon
\begin{equation}
M_F>1.55 {\rm ~~TeV},~~~95\%{\rm ~~C.L.}~,
\label{f3bound}
\end{equation}
where we have considered a critically coupled familon.  This is
consistent with, but somewhat weaker than the $95\%$ C.L. limit
obtained in Ref.\cite{zbounds}, $M_F>1.9$~TeV at critical coupling.

During Run II however, measurements of the dijet 
spectrum at an upgraded Tevatron will yield bounds 
better than those derived from Z-pole observables. 
For instance if we consider the nominal luminosity of $2fb^{-1}$, 
and assume a $30\%$ reduction in the systematic errors, the bound on the 
familon mass for Run~II would be $M_F>2.2$~TeV. An extended 
Tevatron run or the achievement of higher intensities could therefore
result in a mass reach well above that of electroweak precision 
measurements and cover a large fraction of the interesting 
parameter space of this model.


\begin{figure}
\center
\hspace*{0.7cm}
\psfig{figure=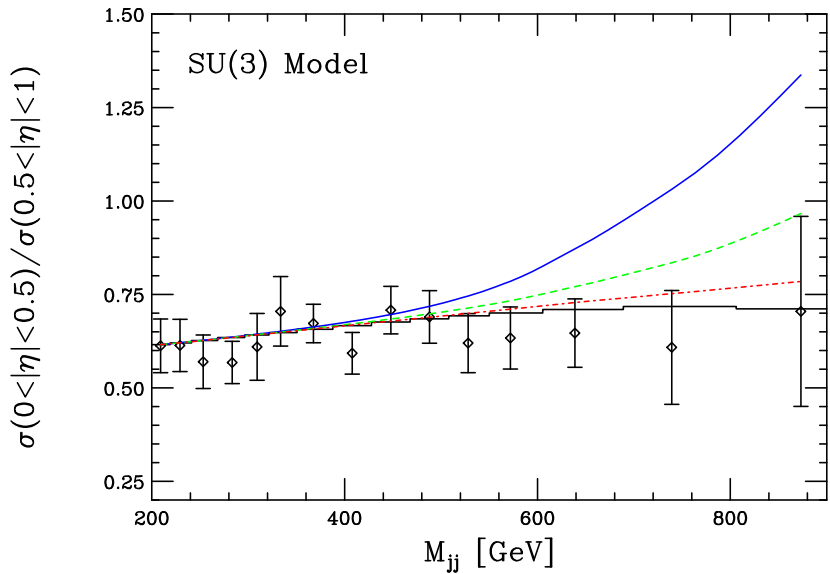,height=3.0in,angle=90}
\caption{\small\em The ratio of cross sections for $(|\eta|<0)/
(0.5<|\eta|1.0)$ vs. the dijet invariant mass, 
for the $SU(3)$ chiral quark family model, for $M_F=1.2~{\rm (solid)},
 1.5~{\rm (dashed)}$ and $2~{\rm TeV~(dot-dash)}$.
The data points are from the D0 measurement \cite{d0kdata}, 
with the error bars including the statistical and systematic errors added 
in quadrature. The histogram is the NLO QCD prediction from JETRAD, using
CTEQ3M parton distribution function.} 
\label{f3vqcd}
\end{figure}

\noindent
In addition to the dijet signal, the chiral quark family model leads
to another potentially interesting signal at hadron colliders:
anomalous single top production.  This occurs due to the existence of
non-diagonal couplings to the family gauge bosons. Although these do
not lead to $|\Delta S|=2$ signals, because of GIM cancellation, there
are flavor changing couplings of quarks. The fact that
the family symmetry commutes with $SU(2)_L$ implies that there will be
tree level familon exchanges such as $d\bar b\to u\bar t$, where
``family number'' is preserved.  The diagrams relevant for single top
production at the Tevatron are s-channel $d\bar b\to u\bar
t$, and t-channel $u \bar d\to t\bar b$ (dominant) and $u\bar b\to t\bar d$. Other
diagrams also are obtained by the replacements $d\to s$ and $u\to c$.
For instance, the s-channel matrix element squared is
\begin{equation}
|{\cal M}(d\bar{b}\rightarrow u\bar{t})|^2 = 
(4\pi)^2 \kappa^2 u(u-m^2_t) \left|{1\over 2}P_s \right|^2 , 
\label{stop}
\end{equation}
Neglecting $m_b$, the t-channel contributions are obtained by
replacing $P_s$ by the $P_t$, where $P_s$ and $P_t$ are the familon
propagators in the corresponding channel as defined in (\ref{pi}).  
\begin{figure}
\center
\hspace*{0.7cm}
\psfig{figure=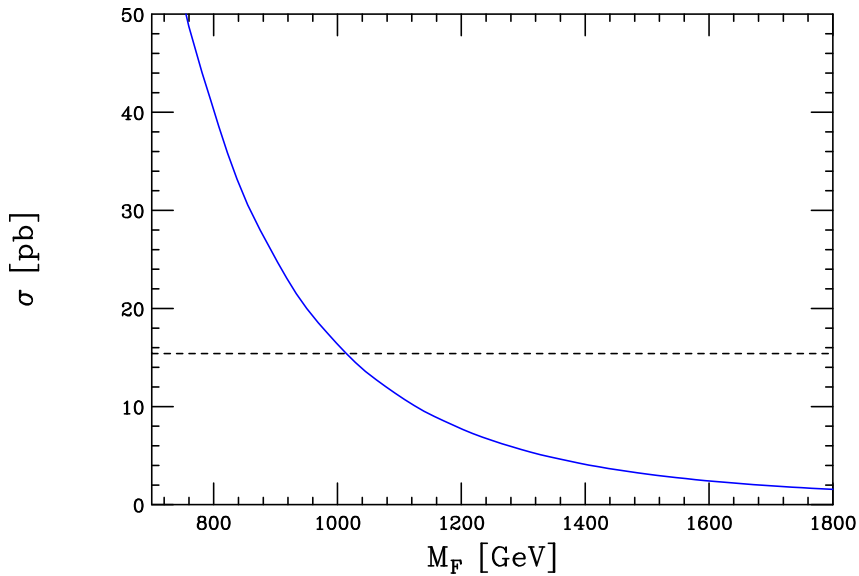,height=3.0in,angle=90}
\caption{\small\em Single Top production cross section in the SU(3)
family model vs. the familon mass, at $\sqrt{s}=2~$TeV. 
The dashed horizontal line corresponds to the 
$95\%$~C.L. bound from Ref.~\cite{cdfst}.
}
\label{str1}
\end{figure}
\noindent
If the coupling is close to critical, these processes will generate
important contributions to the single top production cross section.
In Fig.~\ref{str1} we show the familon induced single top production
cross section at $\sqrt{s}=1.8$~TeV as a function of the familon mass.
The horizontal line is the $95\%$~C.L. upper limit on single top
production as obtained by the CDF
collaboration~\cite{cdfst}.  The most constraining bound,
$\sigma(p\bar p\to tX)<15.4$ pb translates into the familon mass bound
\begin{equation}
M_F>1.02~{\rm TeV}~~~~95\%~{\rm C.L.}.
\end{equation}
This is somewhat weaker than the bound~(\ref{f3bound}) obtained 
from the Run~I dijet data,
but may be improved if a study exploiting the kinematic differences
between the SM and the flavoron signals is undertaken. 

\begin{figure}
\center
\hspace*{0.7cm}
\psfig{figure=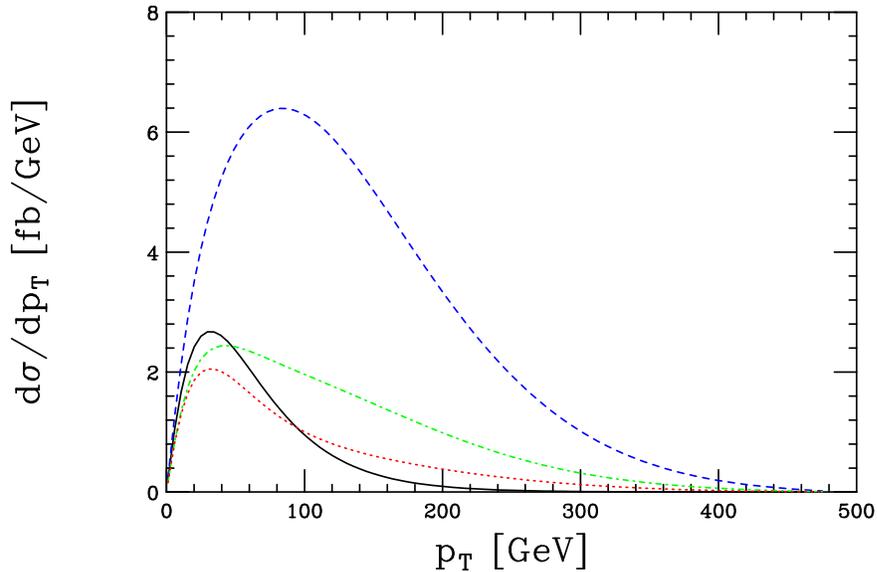,height=3.0in,angle=90}
\caption{\small\em The transverse momentum distribution  
in single top production in the SU(3) family model, for $\sqrt{s}=2~$TeV.
Only the t-channel contribution, leading to the $tb$ final state, is included. 
The solid line is the SM $W^*$ s-channel process. The dashed line corresponds
to $M_F=1.5~$TeV, the dot-dashed line to $M_F=2~$TeV and the dotted line 
to $M_F=2.5~$TeV.  
}
\label{str2}
\end{figure}

In Run~II, the Tevatron will be sensitive to the SM single top
production via $W$-gluon fusion as well as the s-channel $W^*$
exchange. The latter process can be separated from the former by
making use of double b-tagging, since the b quark produced in
association with the top is hard, unlike in $W$-gluon fusion.  In
order to estimate the sensitivity of the Tevatron in Run~II to the
flavoron contribution to single top production, we take only the
dominant flavoron diagram, t-channel mediated $u\bar d  \to t\bar b$.  
We compare this contribution to the s-channel SM assuming these
will be separately observed with the use of double b-tagging~\cite{scott}. 
In Fig.~\ref{str2} we show the $p_T$ distribution of the $b$ quark
produced in association with the top quark for t-channel familon
exchange and s-channel $W^*$ exchange. We see that, for example, for
$M_F=2~$TeV the total ($t\bar b + \bar t b$) cross section is about
$50\%$ larger from familon exchange than in the SM, with the added
feature that the $p_T$ distribution is harder. We conclude that the
sensitivity of Run~II could go as far as $(2-2.5)$~TeV for 2~fb$^{-1}$, or
perhaps higher depending on the sensitivity to be achieved to the SM
s-channel process.  Thus, anomalous single top production could
be the most constraining channel on the $SU(3)$ chiral quark model in
Run~II at the Tevatron.

\subsection{SU(9) Chiral Flavor Symmetry}

We next consider a natural extension of gauging the quark
family symmetry, gauging the full 
SU(9) symmetry of both the color and family multiplicity of the left handed
quarks.  Such a symmetry can be implemented as an extended technicolor 
gauge symmetry (in the spirit of \cite{randall}) or in quark universal 
seesaw models (as in \cite{fuseesaw}). The SU(9) symmetry commutes with
the standard weak $SU(2)_L$ gauge group and acts on the left handed quarks
\beq
Q_L = \left( (t,b)^r, (t,b)^b, (t,b)^g, (c,s)^r,... (u,d)^g \right)_L
\eeq
with $r,g,b$ the three QCD colors. The quark couplings to the 
SU(9) gauge bosons is given by  
\beq
{\cal L} = i g_F B^{a \mu} \bar{Q}_L \Lambda^a \gamma_\mu Q_L~, 
\eeq
with $\Lambda^a$ the generators of SU(9). These include
\beq \label{su9gen}
{1 \over \sqrt{3}} 
\left( \begin{array}{ccc} 
              T^a & 0 & 0 \\ 0 & T^a & 0\\ 0 & 0 & T^a 
       \end{array}\right), 
{1 \over \sqrt{6}} 
\left( \begin{array}{ccc} 
              T^a & 0 & 0 \\ 0 & T^a & 0\\ 0 & 0 & -2 T^a 
       \end{array}\right),
{1 \over \sqrt{2}} 
\left( \begin{array}{ccc} 
              T^a & 0 & 0 \\ 0 & -T^a & 0\\ 0 & 0 & 0  
       \end{array}\right), 
\eeq
where $T^a$ are the 8  3x3 QCD generators. SU(9) further contains
\beq 
{1 \over \sqrt{2}} 
\left( \begin{array}{ccc} 
              0 & T^a & 0 \\ T^a & 0 & 0\\ 0 & 0 & 0 
       \end{array}\right), 
{1 \over \sqrt{12}} 
\left( \begin{array}{ccc} 
              0 & 1 & 0 \\ 1 & 0 & 0\\ 0 & 0 & 0
       \end{array}\right),
{1 \over \sqrt{2}} 
\left( \begin{array}{ccc} 
             0 & -iT^a & 0 \\ iT^a & 0 & 0\\ 0 & 0 & 0  
       \end{array}\right),
{1 \over \sqrt{12}} 
\left( \begin{array}{ccc} 
             0 & -i & 0 \\ i & 0 & 0\\ 0 & 0 & 0  
       \end{array}\right)
\eeq
plus the two other similar sets mixing the remaining families. Finally there
are two diagonal generators
\beq
{1 \over \sqrt{12}} 
\left( \begin{array}{ccc} 
              1 & 0 & 0 \\ 0 & -1 & 0\\ 0 & 0 & 0 
       \end{array}\right), 
{1 \over \sqrt{36}} 
\left( \begin{array}{ccc} 
              1 & 0 & 0 \\ 0 & 1 & 0\\ 0 & 0 & -2
       \end{array}\right)
\eeq

\noindent
The model must also contain interactions which give rise to color for
the right handed quarks. For this reason, we 
include an $SU(3)_{pc}$ proto-color group that acts on the right
handed quarks, which will be combined with the $SU(3)_C$ subgroup of
$SU(9)_L$ to yield ordinary color.  We normalize the proto-color gauge
bosons couplings such that they have the same generators as the SU(9)
bosons
\beq
{\cal L} = {i\over \sqrt{3}} g_{pc} A^{\mu a} \bar{q}_R \gamma_\mu T^a q_R ~.
\eeq
At the flavor breaking scale we assume some massive sector breaks the
$SU(9)_L \times SU(3)_{pc}$ gauge symmetry down to ordinary color
$SU(3)_C$ and a global $SU(3)_F$ group acting on the three families of
quarks. The global $SU(3)_F$ symmetry is sufficient to insure the
absence of tree-level FCNCs \cite{randall}.

For simplicity, we will assume the symmetry breaking sector
has an $SU(9)_L \times SU(9)_{flavor/color}$ chiral flavor symmetry, under
which the symmetry breaking vev transforms as a $(9,\bar{9})$.  The
majority of the SU(9) gauge bosons will then have mass $M_F=g_F V$.
Eight of the $SU(9)_L$ gauge bosons mix with the right handed proto-color group,
giving rise to ordinary color and eight massive gluons. 
The proto-gluons and color-octet flavorons mix through the mass matrix
\beq
(A^\mu, B^\mu) \left( \begin{array}{cc} g_{pc}^2 & -g_{pc} g_F 
\\ -g_{pc} g_F &g_F^2
\end{array} \right)V^2 \left( \begin{array}{c} A_\mu \\ B_\mu \end{array} 
\right)
\eeq
which diagonalizes to
\beq
(X^\mu, G^\mu) \left( \begin{array}{cc} g_{pc}^2 + g_F^2& 0 \\ 0 & 0
\end{array} \right)V^2 \left( \begin{array}{c} X_\mu \\ G_\mu \end{array} \right)
\eeq
where 
\beq
\left( \begin{array}{c} A^\mu \\ B^\mu \end{array} \right) =
\left( \begin{array}{cc} \cos \phi & -\sin \phi  \\ \sin \phi &
\cos \phi
\end{array} \right) \left( \begin{array}{c} G_\mu \\ X_\mu \end{array} \right)
\eeq
with
\beq 
\sin \phi = {g_{pc} \over \sqrt{g_{pc}^2 + g_F^2}}, \hspace{1cm}
\cos \phi = {g_F \over \sqrt{g_{pc}^2 + g_F^2}}~,
\eeq
and $G^\mu$ and $X^\mu$ are the gluon and color-octet flavoron respectively.

\noindent
The low energy QCD coupling, with the standard generator normalization 
is given by
\beq 
g_c = {g_F g_{pc} \over \sqrt{3(g_{pc}^2 + g_F^2)}}
\eeq
which implies that $\kappa_F \geq 3 \alpha_s(2\,{\rm TeV})$.
The interactions of the SM fermions  with the massive color 
octet (with mass $M_{F'}
= \sqrt{g_{pc}^2 + g_F^2} V = M_F/c_\phi$) are given by
\beq
- g_c \tan \phi   X^{a \mu} \bar{q}_R \gamma_\mu T^a q_R +
g_c \cot \phi X^{a \mu} \bar{q}_L \gamma_\mu T^a q_L~,
\eeq
where $\cot \phi = g_F/g_{pc}$.

\noindent
As in the case of $SU(3)_F$, the coupling $g_F$ cannot be too large,
or it would likely induce an EWS breaking condensate
at the flavor scale.  Assuming that at low energies the massive gauge
boson interactions with the SM fermions can be approximated by a NJL
model (ignoring the effects of the mixing of eight of the generators
with proto-color in this estimate), then the critical coupling for
chiral symmetry breaking in that approximation is
\begin{equation}
\kappa_{crit} = {2 N \pi \over (N^2-1)} = 0.71~.
\label{kcf9}
\end{equation}

As was the case in the previous two models, the most 
conspicuous signals are in the dijet spectrum. 
In the Appendix A.3 we list all the relevant matrix
elements for dijet production due to the two gauge bosons with 
masses $M_F$ and $M_F'$. 
\begin{figure}
\center
\hspace*{0.7cm}
\psfig{figure=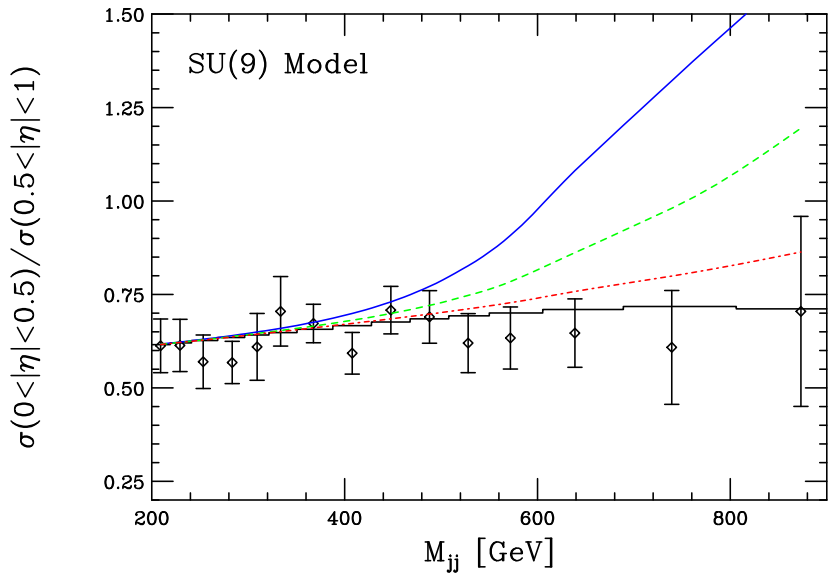,height=3.0in,angle=90}
\caption{\small\em The ratio of cross sections for $(|\eta|<0.5)/
(0.5<|\eta|1.0)$ vs. the dijet invariant mass, 
for the $SU(9)$ chiral flavor model, for $M_F=1.2~{\rm (solid)},
 1.5~{\rm (dashed)}$ and $2~{\rm TeV~(dot-dash)}$.
The data points are from the D0 measurement \cite{d0kdata}, 
with the error bars including the statistical and systematic errors added 
in quadrature. The histogram is the NLO QCD prediction from JETRAD, using
CTEQ3M parton distribution function.} 
\label{f9vqcd}
\end{figure}
In Fig.\ref{f9vqcd} we plot the contributions of these gauge bosons to 
the cross section ratio as a function of their mass, assuming 
for simplicity $M_F=M_F'$. Although, in principle, one could expect 
the effect to be smaller than for the $SU(3)$ chiral familon
due to the fact that the critical coupling in eqn.(\ref{kcf9})
is considerably smaller than that of the $SU(3)$ case, the 
$SU(9)$ flavorons contribute to a large number of diagrams
leading to dijets. In fact, as can be seen in Fig.\ref{f9vqcd},
the effects will be stronger in this model. 
We follow the procedure described earlier to obtain a mass 
constraint from the Tevatron Run~I data. The mass of the $SU(9)$
flavorons is bounded by
\begin{equation}
M_F ~>~~~1.9~ {\rm ~TeV}~, ~~95\% {\rm ~~C.L.} 
\label{f9limit}
\end{equation}
This is very similar to the $95\% {\rm C.L.}$ limit obtained in Ref.
\cite{zbounds} from electroweak precision measurements.  On the other
hand, at $\sqrt{s}=2~$TeV
and with an integrated luminosity of 2~fb$^{-1}$, Run~II at the
Tevatron will put a limit of $M_F~>~~2.7~{\rm TeV}$, 
where we assume a 30\% reduction in systematic errors.  
This covers a large
fraction of the interesting parameter space of this model.

Just as in the chiral quark family model, in the $SU(9)$ model there
are also important contributions to anomalous single top production.
The fact that some of the $SU(9)$ gauge bosons carry color tends to
enhance the interactions when compared to the $SU(3)$ chiral quark
model. On the other hand, the critical coupling in this model is
considerably smaller than that in the $SU(3)$ case, as can be seen
by comparing eqns.~(\ref{kcf9}) with (\ref{kcsu3}).  The net effect is
a reduction in the single top signal shown in Fig~\ref{str2}, by a
factor of
\begin{equation} 
\left(\frac{\kappa_F^{SU(9)}}{\kappa_F^{SU(3)}}\right)^2
\times\left({14\over 9}\right)\simeq 0.15
\label{redfac}
\end{equation} 
at critical coupling.
Since the cross section falls approximately as $1/M_F^4$, 
this will result in a familon mass bound that is smaller than the one to be obtained 
in the $SU(3)$ model by a factor of about $\sqrt[4]{0.15}\simeq 0.60$.
Thus, since our expectations for Run~II in the single top channel
in the $SU(3)$ model put the reach somewhere around $M_F>(2-2.5)~$TeV, we conclude 
that the reach of this channel for the $SU(9)$ flavoron is still below the 
Run~I mass limit eqn.~(\ref{f9limit}) that we extracted from the dijet data. 
Although more detailed studies of the single top signal (for instance including all 
possible single top final states) are possible, we can safely conclude that this
channel will not be competitive with the dijet signal in the $SU(9)$ model at
the Tevatron. 


\subsection{SU(12) Chiral Flavor Symmetry}

The final model we consider is one in which we gauge the
full SU(12) flavor symmetry of all the left handed SM fermion
doublets \cite{randall,fuseesaw}
\beq
Q_L = \left( (t,b)^r, (t,b)^b, (t,b)^g, (\nu_\tau, \tau), 
(c,s)^r,... (\nu_e,e) 
\right)_L~.
\eeq
This  is similar to the $SU(9)$ model, but it also includes 
a proto-hypercharge interaction that, after the $SU(12)$ breaking, 
gives rise to the SM $U(1)_Y$. 
The flavor gauge interactions act as
\begin{equation}
{\cal L} = i g_F B^{a \mu} \bar{Q}_L \Lambda^a \gamma_\mu Q_L~,
\label{lf}
\end{equation}
with $\Lambda^a$ the generators of SU(12), which may be conveniently  
broken down into the following groupings
\beq \label{su12gen}
{1 \over \sqrt{3}} 
\left( \begin{array}{ccc} 
              P^a & 0 & 0 \\ 0 & P^a & 0\\ 0 & 0 & P^a 
       \end{array}\right), 
{1 \over \sqrt{6}} 
\left( \begin{array}{ccc} 
              P^a & 0 & 0 \\ 0 & P^a & 0\\ 0 & 0 & -2 P^a 
       \end{array}\right),
{1 \over \sqrt{2}} 
\left( \begin{array}{ccc} 
              P^a & 0 & 0 \\ 0 & -P^a & 0\\ 0 & 0 & 0  
       \end{array}\right)
\eeq
where $P^a$ are the 15 4x4 Pati-Salam generators consisting of
8 3x3 blocks that are QCD, 6 step operators between the quarks and leptons
and the diagonal generator $1/\sqrt{24}$ diag$(1,1,1,-3)$. SU(12) further contains
\beq 
{1 \over \sqrt{2}} 
\left( \begin{array}{ccc} 
              0 & P^a & 0 \\ P^a & 0 & 0\\ 0 & 0 & 0 
       \end{array}\right), 
{1 \over \sqrt{16}} 
\left( \begin{array}{ccc} 
              0 & 1 & 0 \\ 1 & 0 & 0\\ 0 & 0 & 0
       \end{array}\right),
{1 \over \sqrt{2}} 
\left( \begin{array}{ccc} 
             0 & -iP^a & 0 \\ iP^a & 0 & 0\\ 0 & 0 & 0  
       \end{array}\right),
{1 \over \sqrt{16}} 
\left( \begin{array}{ccc} 
             0 & -i & 0 \\ i & 0 & 0\\ 0 & 0 & 0  
       \end{array}\right)
\eeq
plus the two other similar sets mixing the remaining families. Finally there
are two diagonal generators
\beq
{1 \over \sqrt{16}} 
\left( \begin{array}{ccc} 
              1 & 0 & 0 \\ 0 & -1 & 0\\ 0 & 0 & 0 
       \end{array}\right), 
{1 \over \sqrt{48}} 
\left( \begin{array}{ccc} 
              1 & 0 & 0 \\ 0 & 1 & 0\\ 0 & 0 & -2
       \end{array}\right)
\eeq
\noindent
In order to ensure the SM gauge groups emerge at low energies we must 
again introduce
a proto-color group as in the SU(9) model above. The first 8 generators of
SU(12) in (\ref{su12gen}) are the same as those in the SU(9) model 
(\ref{su9gen}) and hence the discussion of the mixing between the proto-color
and the 8 SU(12) gauge bosons follows the discussion in the SU(9) model 
exactly.
\noindent
In addition, in the SU(12) model we must also include a proto-hypercolor gauge
boson. Since the Pati-Salam diagonal generator in the first set of generators in 
(\ref{su12gen}) is the traditional generator for the hypercharge boson's 
coupling to left handed fermions, the  proto-hypercharge gauge boson
only has to couple to the right handed fermions. 
The result of the mixing of these two gauge bosons is the massless SM hypercharge 
gauge boson plus a massive gauge boson coupling to both left and right handed 
fermions. 

\noindent
If the interactions of flavoron gauge bosons in
Eq.~(\ref{lf}) at low energies can be modeled by a NJL lagrangian with
coupling $4 \pi \kappa / 2! M_{F}^2$, the critical coupling for chiral
symmetry breaking is calculated to be
\begin{equation}
\kappa_{crit} = {2 N \pi \over (N^2-1)} = 0.53~,
\label{kcsu12}
\end{equation}
somewhat smaller than in the $SU(9)$ case in Eq.~(\ref{kcf9}).
Note that combined with the lower constraint from the ability to reproduce 
the QCD coupling ($\kappa_F \geq 3 \alpha_s(2\,{\rm TeV}) \simeq 0.3$) 
there is a relatively small window of allowed couplings.

\noindent
Although this model results in various signals at the Tevatron ---
such as quark scatterings similar to those of the $SU(9)$ model as
well as anomalous contributions to Drell-Yan production arising from
the flavoron couplings to leptons --- the energy scale of this
scenario is severely constrained by data from experiments of Atomic
Parity Violation (APV) in Cesium.  The parity-violating part of the
electron-nucleon interaction can be written as
\begin{equation}
{\cal L}_{eq}=\frac{G_F}{\sqrt{2}}\sum_{q=u,d}\{
C_{1q}(\bar e\gamma_\mu\gamma_5 e)(\bar q\gamma^\mu q) 
+ 
C_{2q}(\bar e\gamma_\mu e)(\bar q\gamma^\mu\gamma_5 q)\}~,
\label{pvl}
\end{equation}
where the coefficients $C_{1q}$ and $C_{2q}$ are given in the SM by
\begin{equation}
C_{1q}^{SM}=-(T_3^q -2Q_q\sin^2\theta)~~~~~, 
C_{2q}^{SM}=-T_3^q(1-4\sin^2\theta)~,
\end{equation}
and $T_3^q$ is the third component of the quark isospin. 
The atomic weak charge is then defined as
\begin{equation}
Q_W=-2\{C_{1u}(2Z+N) + C_{1d}(N+2Z)\}~,
\end{equation} 
with $Z$ and $N$ the number of protons and neutrons respectively. 
The APV experiment finds the atomic charge of Cesium to be 
\cite{apvexp} $Q_W=-72.06\pm 0.28\pm 0.34 $, whereas the 
SM prediction \cite{apvth} is  $Q_W=-73.09\pm 0.03$.
This translates into a deviation from the SM prediction of 
\begin{equation}
\Delta Q_W=1.33\pm 0.44~.
\label{expdev}
\end{equation}
We can write the deviations of $Q_W$ as 
\begin{equation}
\Delta Q_W=-376\Delta C_{1u} -422\Delta C_{1d}~.
\label{dqw}
\end{equation} 

\noindent
The SU(12) model gives rise to various contributions to $\Delta Q_W$. 
However, by far the largest of these corresponds to a step operator 
from the generators in Eq.~(\ref{su12gen}) which connect quarks to leptons. 
These result in the non-diagonal effective coupling 
\begin{equation}
-\frac{g_F^2}{8\,M_F^2} (\bar e_L\gamma_\mu d_L)(\bar d_L\gamma^\mu e_L)~. 
\label{etall}
\end{equation}
After Fierzing and decomposing into the proper vector and axial pieces, 
the contribution in (\ref{etall}) gives rise to an effect in the 
weak charge of Cesium given by
\begin{equation}
\Delta Q_W^{F} = -80.4\kappa_F\frac{(1\,{\rm TeV})^2}{M_F^2} =  
-42.6\frac{(1\,{\rm TeV})^2}{M_F^2}~,
\label{apv12}
\end{equation}
where $M_F$ is understood to be measured in TeV, and the last equality
is obtained by using $\kappa_F=\kappa_{\rm crit.}$ as defined in
(\ref{kcsu12}).  Thus not only is this a {\it large} contribution to
$Q_W(Cs)$, but it also has the opposite sign of Eq.~(\ref{expdev}).
For instance, the $3\sigma$ bound would be $M_F>12$~TeV.  More
conservatively, we can estimate the sensitivity of the APV measurement
by taking the error in Eq.~(\ref{expdev}) as the possible size of the
effect.  This translates into $M_F>9.8~$TeV.  From the model building
point of view this is an undesirably large mass scale and raises the
issue of fine-tuning.  In any event, it is clear that the APV
experiment forces the mass scale in the $SU(12)$ model to be very high
and out of reach of the Tevatron.

Finally, we point out that the constraint on the $SU(12)$ model
resulting from Eq.~(\ref{apv12}) is more general since it cannot be
completely evaded by lowering the coupling below its critical value.
As we mentioned earlier, in order to obtain the correct QCD coupling,
$\kappa_F$ must satisfy $\kappa_F\geq 3 \alpha_s(2\,{\rm TeV})$. Then,
its minimum value of approximately $0.3$ translates into the
bound $M_F > 7.4~$TeV.

\section{Conclusions}

We have studied the Tevatron collider bounds on two models of
broken, gauged,  chiral flavor symmetries; an SU(3) chiral family symmetry 
and an SU(9) chiral flavor symmetry 
of the SM quarks. These symmetries have been
proposed as playing a significant role in theories of EWS breaking
and fermion mass generation and are blessed with a GIM mechanism that 
suppresses FCNCs allowing the gauge bosons to be relatively light. 
The strongest
Tevatron signals result in dijet production and single top production.
We summarize the current limits, from precision data \cite{zbounds} and Run I,
on the critically coupled gauge boson masses in Table 1 - 
they are comparable. The Run II expectations are also displayed
and should become the leading constraints on the models.

In the $SU(3)$ model both, dijet and anomalous single top production, are likely
to be important signals. On the other hand, in the 
$SU(9)$ model the dijet cross
section receives a large enhancement due to the fact that some of the 
flavor gauge bosons carry color, resulting in more diagrams contributing (see 
Appendix~A.3). However, 
since the critical coupling is considerably smaller than in the 
$SU(3)$ case, the single top signal -- even after taking into account the color 
enhancement -- is reduced. Thus, the single top channel is crucial in order 
to separate these two models as the possible origin of a hypothetical deviation in the 
dijet sample.

For comparison we also display in Table~I the equivalent limits for the Universal Coloron 
model of \cite{simmons, bertram} - 
in this model the chiral $SU(3)_L \times SU(3)_R$ color
group of the quarks is gauged and broken to the QCD group leaving 
axially coupling massive colorons. This
model is considerably more strongly constrained in part because of its 
large critical coupling and because the dijet channel is a particularly
good probe of extra color like interactions. 
It is notable that 
in the models we have explored the gauge bosons are potentially lighter, as
one might hope if they played a role in EWS breaking, and that the Tevatron
can hope to probe interesting regions of parameter space.

Finally we have pointed out a further low energy precision constraint on
models where the flavor symmetry is enlarged to include the lepton sector.
In particular an SU(12) gauged chiral flavor model 
gives contributions in low energy 
atomic parity violation experiments that place the bound on the gauge boson
masses out of the Tevatron's reach.

\phantom{xxxx}\vspace{0.1in}
\begin{center}
\begin{tabular}{|c||c|c|c|}
\hline
 & EPM& Run~I&Run~II  \\ \hline\hline
  Universal Coloron & 3 & 4.3 & 7 \\ \hline 
$SU(3)_F$ & 1.9 & 1.55 & 2.5 (single top) \\ \hline
$SU(9)_F$ & 1.9 & 1.9 & 2.7 \\ \hline
$SU(12)_F$ & 10 (APV) & No reach & No reach \\ \hline
\end{tabular}
\begin{flushleft}
Table~I:~{\small\em 
The $95\%~$ C.L. bounds (or sensitivity) on the models discussed.
The numbers correspond to the mass of the gauge bosons in TeV
if its coupling is critical. 
The first column comes from electroweak
precision measurements and is taken from Ref.~\cite{zbounds}.
The Run~I bounds as well as the Run~II sensitivities (for $2fb^{-1}$)
summarize our results. 
They come from the dijet analysis, with the exception of the Run~II reach for the 
$SU(3)$ chiral quark model which comes from single top production.
}
\end{flushleft}
\end{center}
\phantom{xxxx}\vspace{0.1in}


\centerline{\bf Acknowledgments}

{\em This work was supported in part by the Department of Energy under
  grants DE-FG02-91ER40676 and DE-FG02-95ER40896. 
  N.E. is grateful for the support of a PPARC Advanced Research Fellowship.
  G.B. acknowledges the hospitality of the High Energy Physics Group
  at the University of Sao Paulo, where part of this work was completed.}



\appendix

\section*{Appendix: Cross Sections}
\setcounter{equation}{0}
\renewcommand{\theequation}{A.\arabic{equation}}

We present some standard tree-level expressions for cross sections. 
\beq
{d \sigma \over d t} = {1 \over 16 \pi}{1\over s^2} |{\cal M}|^2
\label{dsdt}
\eeq
To obtain the full cross section we must average over initial states and 
sum over final states. Summing over spins and splitting the matrix 
element into chiral components we have
\begin{eqnarray}
\bar{L}L \rightarrow \bar{L} L: & {1 \over 4} \sum_{spin} |{\cal M}|^2 = & 
u^2|\sum_i P_iQ_i|^2\\
\bar{L}R \rightarrow \bar{L} R: & {1 \over 4} \sum_{spin} |{\cal M}|^2 = & 
s^2|\sum_i P_iQ_i|^2\\
\bar{L}L \rightarrow \bar{R} R: & {1 \over 4} \sum_{spin} |{\cal M}|^2 = & 
t^2|\sum_i P_iQ_i|^2
\end{eqnarray}
where $P_i$ is the propagator factor associated with each diagram taking
the form
\beq
P_i = {-i \over q_i^2 - M_F^2 + i \Gamma_F M_F}
\label{pi}
\eeq
and one must sum over all gauge bosons and $q_i^2=s,t$ channels. $Q_i$ 
are the group theory factors associated with each diagram.
Application of the above construction kit and averaging over initial
color states (1/9) and summing final color states gives the QCD
backgrounds and flavor model contributions to dijet processes.

\subsection*{A.1~~~QCD Backgrounds}

\begin{eqnarray}
{d \sigma \over d t} (q q \rightarrow qq) & = &
{4 \pi \alpha_s^2 \over 9 s^2}\left( {u^2 + s^2 \over t^2}
+ {t^2+s^2\over u^2} - {2 \over 3}{s^2\over ut}\right)\\
{d \sigma \over d t} (q \tilde{q} \rightarrow q\tilde{q}) & = &
{4 \pi \alpha_s^2 \over 9 s^2}\left({s^2+u^2 \over t^2} \right)\\
{d \sigma \over d t} (q\bar{q} \rightarrow \tilde{q}\bar{\tilde{q}}) & = &
{4 \pi \alpha_s^2 \over 9 s^4}(t^2+u^2)\\
{d \sigma \over d t} (q\bar{q} \rightarrow q\bar{q}) & = &
{4 \pi \alpha_s^2 \over 9 s^2}\left( {s^2+u^2\over t^2} + {t^2+u^2 \over s^2}
- {2\over 3} {u^2\over st}\right)\\
{d \sigma \over d t} (q\bar{\tilde{q}} \rightarrow q\bar{\tilde{q}}) & = &
{4 \pi \alpha_s^2 \over 9 s^s}\left({s^2+u^2 \over t^2}\right)\\
{d \sigma \over d t} (gg \rightarrow q\bar{q}) & = &
{\pi \alpha_s^2 \over 6 s^2} \left( {u \over t} + {t \over u} 
- {9 \over 4}{t^2+u^2 \over s^2}\right)\\
{d \sigma \over d t} (q\bar{q} \rightarrow gg) & = &
{32 \pi \alpha_s^2 \over 27 s^2} \left( {u \over t} + {t \over u} 
- {9 \over 4}{t^2+u^2 \over s^2}\right)\\
{d \sigma \over d t} (qg \rightarrow qg) & = &
{4 \pi \alpha_s^2 \over 9 s^2}\left(- {u\over s} - {s\over u} + {9\over 4} 
{s^2+u^2\over t^2} \right)\\
{d \sigma \over d t} (gg \rightarrow gg) & = &
{9 \pi \alpha_s^2 \over 2 s^2} \left( 3 - {tu\over s^2} - {su\over t^2} 
- {st \over u^2} \right)
\end{eqnarray}

\subsection*{A.2~~~Chiral Quark Family Symmetry: Matrix Elements
into dijets.}

\begin{eqnarray}
\Delta |{\cal M}(qq\rightarrow qq)|^2 &  
= & (4\pi)^2 \kappa^2 s^2 
\left|{1 \over 3} P_t - {1 \over 3} P_u \right|^2\nonumber\\ 
& & - {(4\pi)^2 \kappa \alpha_s s^2\over 9} Re \left( {1 \over t} P_t +
{1 \over u} P_u - {1\over u}P_t - {1 \over t} P_u \right)\\
\Delta |{\cal M}(ud\rightarrow ud)|^2 &  
= & {(4\pi)^2 \kappa^2 s^2\over 9} |P_t|^2 
 + {(4\pi)^2 \kappa \alpha_s s^2\over 9} Re \left( {1 \over t} P_t \right)\\
\Delta |{\cal M}(us\rightarrow us)|^2 &  
= & {(4\pi)^2 \kappa^2 s^2\over 36} |P_t|^2 
 + {(4\pi)^2 \kappa \alpha_s s^2\over 18} Re \left( {1 \over t} P_t \right)\\
\Delta |{\cal M}(ds\rightarrow ds)|^2 &  
= & (4\pi)^2 \kappa^2 s^2 \left| {1 \over 6}P_t + {1 \over 2} P_u\right|^2 
 + {(4\pi)^2 \kappa \alpha_s s^2\over 3} Re \left( {1 \over 6 t} P_t + {1 \over 2 t} P_u\right)\\
\Delta |{\cal M}(q\bar{q}\rightarrow q\bar{q})|^2 &  
= & (4\pi)^2 \kappa^2 u^2
\left|{1 \over 3} P_t - {1 \over 3} P_s \right|^2 
 - {(4\pi)^2 \kappa \alpha_s u^2\over 9} Re \left( {1 \over s} P_t +
{1 \over t} P_s \right)\\
\Delta |{\cal M}(u\bar{u}\rightarrow d\bar{d})|^2 &  
= & (4\pi)^2 \kappa^2 u^2
\left |{1\over 3} P_s \right|^2\\
\Delta |{\cal M}(u\bar{u}\rightarrow s\bar{s})|^2 &  
= & (4\pi)^2 \kappa^2 u^2
\left |{1\over 6} P_s \right|^2\\
\Delta |{\cal M}(d\bar{d}\rightarrow s\bar{s})|^2 &  
= & (4\pi)^2 \kappa^2 u^2
\left |{1\over 6} P_s + {1 \over 2} P_t\right|^2
- {(4\pi)^2 \kappa \alpha_s u^2\over 6} Re\left( {1\over s}P_t \right)\\
\Delta |{\cal M}(s\bar{d}\rightarrow s\bar{d})|^2 &  
= & (4\pi)^2 \kappa^2 u^2 \left|{1\over 2} P_s 
+ {1\over 6} P_t \right|^2 
- {(4\pi)^2 \kappa \alpha_s u^2\over 6} Re\left({1 \over t} P_s\right)\\
\Delta |{\cal M}(u\bar{d}\rightarrow u\bar{d})|^2 & 
= & (4\pi)^2 \kappa^2 u^2 \left|{1\over 3} P_t \right|^2\\
\Delta |{\cal M}(u\bar{s}\rightarrow u\bar{s})|^2 & 
= & (4\pi)^2 \kappa^2 u^2 \left|{1\over 6} P_t \right|^2~,\\
\end{eqnarray}
where $P_s$, $P_t$ and $P_u$ are defined by eqn.(A.5) and 
basically reflect the gauge boson propagator in the appropriate channel. 
Among the familon contributions we also include the interference 
with the gluon.

\subsection*{A.3~~~SU(9) Chiral Flavor Symmetry. Matrix elements into
dijets.}

\begin{eqnarray}
|{\cal M}(q_L q_L\rightarrow q_L q_L)|^2 &  
= & 
{2 (4 \pi)^2 s^2  \over 9} \left(
\left|  {\alpha_s \over t} + {2 \kappa\over 3} P^F_t +\alpha_s \cot^2 \phi 
P_t^{F'}\right|^2 \right. \nonumber\\
&& \left.
+ \left|  {\alpha_s \over u}+ {2\kappa\over 3} P^F_u +\alpha_s \cot^2 \phi 
P_u^{F'} \right|^2 \right.\\
&& \nonumber
\left. -{2 \over 3} Re\left[ ({\alpha_s \over t} + {2\kappa\over 3} P^F_t
+ \alpha_s \cot^2 \phi P_t^{F'})
({\alpha_s \over u}+ {2\kappa\over 3} P^F_u+ \alpha_s \cot^2 \phi
P_u^{F'}) \right]\right) \\
|{\cal M}(q_R q_R\rightarrow q_R q_R)|^2 &  
= & 
{2 (4 \pi)^2  s^2 \over 9}  \left(
\left|  {\alpha_s \over t} + \alpha_s \tan^2 \phi 
P_t^{F'} \right|^2 
+ \left|  {\alpha_s \over u} +  \alpha_s \tan^2 \phi 
P_u^{F'} \right|^2 \right. \nonumber \\
&& \left.
-{2 \over 3} Re\left[ ({\alpha_s \over u}+\alpha_s \tan^2 \phi 
P_u^{F'})( {\alpha_s \over t} +  \alpha_s \tan^2 \phi 
P_t^{F'})\right]\right)\\
|{\cal M}(q_Lq_R\rightarrow q_Lq_R)|^2 &  
= & 
{2 (4 \pi)^2 u^2 \over 9} \left| {\alpha_s \over t} -  \alpha_s P_t^{F'} 
\right|^2 \\
|{\cal M}(q_Lq_R\rightarrow q_Rq_L)|^2 &  
= & 
{2 (4 \pi)^2  t^2 \over 9} \left| {\alpha_s \over u} 
- \alpha_s P^{F'}_u \right|^2 
\end{eqnarray}

\begin{eqnarray}
|{\cal M}(u_L d_L\rightarrow u_L d_L)|^2 &  
= & 
{2 (4 \pi)^2 s^2  \over 9} 
\left|  {\alpha_s \over t} + {2 \kappa\over 3} P^F_t + \alpha_s \cot^2 
\phi P_t^{F'}\right|^2 \\
|{\cal M}(u_R d_R\rightarrow u_R d_R)|^2 &  
= & 
{2 (4 \pi)^2  s^2 \over 9}  
\left|  {\alpha_s \over t} +  \alpha_s \tan^2 \phi 
P_t^{F'} \right|^2\\
|{\cal M}(u_Ld_R\rightarrow u_Ld_R)|^2 &  
= & |{\cal M}(u_Rd_L\rightarrow u_Rd_L)|^2 =
{2 (4 \pi)^2 u^2 \over 9} \left| {\alpha_s \over t} - \alpha_s P_t^{F'} 
\right|^2 
\end{eqnarray}

\begin{eqnarray}
|{\cal M}(u_L s_L\rightarrow u_L s_L)|^2 &  
= & 
{2 (4 \pi)^2 s^2  \over 9} 
\left|  {\alpha_s \over t} - { \kappa\over 3} P^F_t + \alpha_s \cot^2 
\phi P_t^{F'}\right|^2 \\
|{\cal M}(u_R s_R\rightarrow u_R s_R)|^2 &  
= & 
{2 (4 \pi)^2  s^2 \over 9} 
\left|  {\alpha_s \over t} +  \alpha_s \tan^2 \phi 
P_t^{F'} \right|^2\\
|{\cal M}(u_Ls_R\rightarrow u_Ls_R)|^2 &  
= & |{\cal M}(u_Rs_L\rightarrow u_Rs_L)|^2 =
{2 (4 \pi)^2 u^2 \over 9} \left| {\alpha_s \over t} -  \alpha_s P_t^{F'} 
\right|^2 
\end{eqnarray}

\begin{eqnarray}
|{\cal M}(d_L s_L\rightarrow d_L s_L)|^2 &  
= & 
{2 (4 \pi)^2 s^2  \over 9} \left(
\left|  {\alpha_s \over t} - { \kappa\over 3} P^F_t + \alpha_s \cot^2 \phi 
P_t^{F'}\right|^2 \right. \nonumber \\
&& \left. + |\kappa P_s^F|^2 -{2\over 3} Re\left[ \kappa P_s^F
({\alpha_s \over t} - { \kappa\over 3} P^F_t + \alpha_s \cot^2 \phi 
P_t^{F'})\right]\right)\\
|{\cal M}(d_R s_R\rightarrow d_R s_R)|^2 &  
= & 
{2 (4 \pi)^2  s^2 \over 9}  
\left|  {\alpha_s \over t} +\alpha_s \tan^2 \phi 
P_t^{F'} \right|^2\\
|{\cal M}(d_Ls_R\rightarrow d_Ls_R)|^2 &  
= & |{\cal M}(d_Rs_L\rightarrow d_Rs_L)|^2 =
{2 (4 \pi)^2 u^2 \over 9} \left| {\alpha_s \over t} -  \alpha_s P_t^{F'} 
\right|^2 
\end{eqnarray}

\begin{eqnarray}
|{\cal M}(q_L\bar{q_L}\rightarrow q_L\bar{q_L})|^2 &  
= & 
{2 (4 \pi)^2 u^2  \over 9} \left(
\left|  {\alpha_s \over s} + {2 \kappa\over 3} P^F_s +\alpha_s \cot^2 \phi 
P_s^{F'}\right|^2 \right. \nonumber \\
&& \left.
+ \left|  {\alpha_s \over t}+ {2\kappa\over 3} P^F_t + \alpha_s \cot^2 \phi 
P_t^{F'} \right|^2 \right.\\
&& \nonumber
\left. -{2 \over 3} Re\left[ ({\alpha_s \over s} + {2\kappa\over 3} P^F_s
+\alpha_s \cot^2 \phi P_s^{F'})
({\alpha_s \over t}+ {2\kappa\over 3} P^F_t+ \alpha_s \cot^2 \phi   
P_t^{F'}) \right]\right) \\
|{\cal M}(q_R\bar{q_R}\rightarrow q_R\bar{q_R})|^2 &  
= & 
{2 (4 \pi)^2  u^2 \over 9}  \left(
\left|  {\alpha_s \over s} +  \alpha_s \tan^2 \phi 
P_s^{F'} \right|^2 
+ \left|  {\alpha_s \over t} +  \alpha_s \tan^2 \phi 
P_t^{F'} \right|^2 \right. \\
&& \left.
-{2 \over 3} Re\left[ ({\alpha_s \over s}+  \alpha_s \tan^2 \phi 
P_s^{F'})( {\alpha_s \over t} + \alpha_s \tan^2 \phi 
P_t^{F'})\right]\right)\\
|{\cal M}(q_L\bar{q_L}\rightarrow q_R\bar{q_R})|^2 &  
= & |{\cal M}(q_R\bar{q_R}\rightarrow q_L\bar{q_L})|^2
={2 (4 \pi)^2 t^2 \over 9} \left| {\alpha_s \over s} -  \alpha_s P_s^{F'} 
\right|^2 \\
|{\cal M}(q_L\bar{q_R}\rightarrow q_L\bar{q_R})|^2 &  
= & |{\cal M}(q_R\bar{q_L}\rightarrow q_R\bar{q_L})|^2=
{2 (4 \pi)^2  s^2 \over 9} \left| {\alpha_s \over t} 
-\alpha_s P^{F'}_t\right|^2 
\end{eqnarray}

\begin{eqnarray}
|{\cal M}(u_L\bar{u_L}\rightarrow d_L\bar{d_L})|^2 &  
= & 
{2 (4 \pi)^2 u^2  \over 9} 
\left|  {\alpha_s \over s} + {2 \kappa \over 3}P_s^{F} 
+ \alpha_s \cot^2 \phi  P_s^{F'} \right|^2\\
|{\cal M}(u_R\bar{u_R}\rightarrow d_R\bar{d_R})|^2 &  
= & 
{2 (4 \pi)^2  u^2 \over 9}  
\left|  {\alpha_s \over s} + \alpha_s \cot^2 \phi  P_s^{F'} \right|^2 
\\
|{\cal M}(u_L\bar{u_L}\rightarrow d_R\bar{d_R})|^2 &  
= & |{\cal M}(u_R\bar{u_R}\rightarrow d_L\bar{d_L})|^2 =
{2 (4 \pi)^2 t^2 \over 9} \left| {\alpha_s \over s} - \alpha_s P_s^{F'} 
\right|^2 
\end{eqnarray}

\begin{eqnarray}
|{\cal M}(u_L\bar{u_L}\rightarrow s_L\bar{s_L})|^2 &  
= & 
{2 (4 \pi)^2 u^2  \over 9} 
\left|  {\alpha_s \over s} - { \kappa \over 3}P_s^{F} 
+ \alpha_s \cot^2 \phi P_s^{F'} \right|^2\\
|{\cal M}(u_R\bar{u_R}\rightarrow s_R\bar{s_R})|^2 &  
= & 
{2 (4 \pi)^2  u^2 \over 9}  
\left|  {\alpha_s \over s} + \alpha_s \tan^2 \phi   P_s^{F'} \right|^2 
\\
|{\cal M}(u_L\bar{u_L}\rightarrow s_R\bar{s_R})|^2 &  
= & |{\cal M}(u_R\bar{u_R}\rightarrow s_L\bar{s_L})|^2 =
{2 (4 \pi)^2 t^2 \over 9} \left| {\alpha_s \over s} -  \alpha_s P_s^{F'} 
\right|^2 
\end{eqnarray}

\begin{eqnarray}
|{\cal M}(d_L\bar{d_L}\rightarrow s_L\bar{s_L})|^2 &  
= & 
{2 (4 \pi)^2 u^2  \over 9} \left(
\left|  {\alpha_s \over s} - { \kappa \over 3}P_s^{F} 
+ \alpha_s \cot^2 \phi P_s^{F'} \right|^2 
+ {1 \over 2} \left|  \kappa P_t^F  \right|^2\right. \\
&& \left.
-{1 \over 3} Re\left[ ({\alpha_s \over s} - { \kappa \over 3}P_s^{F} 
+ \alpha_s \cot^2 \phi P_s^{F'})\kappa P_t^F \right]\right) \\
|{\cal M}(d_R\bar{d_R}\rightarrow s_R\bar{s_R})|^2 &  
= &  {2 (4 \pi)^2  u^2 \over 9}  
\left|  {\alpha_s \over s} + \alpha_s \tan^2 \phi P_s^{F'} \right|^2 
\\
|{\cal M}(d_L\bar{d_L}\rightarrow s_R\bar{s_R})|^2 &  
= & |{\cal M}(d_R\bar{d_R}\rightarrow s_L\bar{s_L})|^2 =
{2 (4 \pi)^2 t^2 \over 9} \left| {\alpha_s \over s} - \alpha_s P_s^{F'} 
\right|^2 
\end{eqnarray}

\begin{eqnarray}
|{\cal M}(s_L\bar{d_L}\rightarrow s_L\bar{d_L})|^2 &  
= & 
{2 (4 \pi)^2 u^2  \over 9} \left(
\left|  {\alpha_s \over t} - {\kappa \over 3}P_t^F + \alpha_s \cot^2 \phi
P_t^{F'}  \right|^2
+ {1 \over 2} \left|  \kappa P_s^F  \right|^2 \right. \nonumber\\
&& \left.
-{1 \over 3} Re\left[\kappa P_s^F ({\alpha_s \over t} 
- {\kappa \over 3}P_t^F + \alpha_s \cot^2 \phi P_t^{F'}) \right]\right) \\
|{\cal M}(s_R\bar{d_R}\rightarrow s_R\bar{d_R})|^2 &  
= & 
{2 (4 \pi)^2  u^2 \over 9}  
\left|  {\alpha_s \over t} + \alpha_s \tan^2 \phi P_t^{F'} 
\right|^2 \\
|{\cal M}(s_L\bar{d_R}\rightarrow s_L\bar{d_R})|^2 &  
= & |{\cal M}(s_R\bar{d_L}\rightarrow s_R\bar{d_L})|^2 =
{2 (4 \pi)^2 s^2 \over 9} \left| {\alpha_s \over t} -  \alpha_s P_t^{F'} 
\right|^2 
\end{eqnarray}

\begin{eqnarray}
|{\cal M}(u_L\bar{d_L}\rightarrow u_L\bar{d_L})|^2 &  
= & 
{2 (4 \pi)^2 u^2  \over 9} \left(
\left|  {\alpha_s \over t} + {2 \kappa \over 3}P_t^F + \alpha_s \cot^2 \phi
P_t^{F'}  \right|^2 \right) \\
|{\cal M}(u_R\bar{d_R}\rightarrow u_R\bar{d_R})|^2 &  
= & 
{2 (4 \pi)^2  u^2 \over 9}  
\left|  {\alpha_s \over t} + \alpha_s \tan^2 \phi P_t^{F'} 
\right|^2 \\
|{\cal M}(u_L\bar{d_R}\rightarrow u_L\bar{d_R})|^2 &  
= & |{\cal M}(u_r\bar{d_L}\rightarrow u_R\bar{d_L})|^2 =
{2 (4 \pi)^2 s^2 \over 9} \left| {\alpha_s \over t} -  \alpha_s P_t^{F'} 
\right|^2 
\end{eqnarray}

\begin{eqnarray}
|{\cal M}(u_L\bar{s_L}\rightarrow u_L\bar{s_L})|^2 &  
= & 
{2 (4 \pi)^2 u^2  \over 9} \left(
\left|  {\alpha_s \over t} - { \kappa \over 3}P_t^F + \alpha_s \cot^2 \phi
P_t^{F'}  \right|^2 \right) \\
|{\cal M}(u_R\bar{s_R}\rightarrow u_R\bar{s_R})|^2 &  
= & 
{2 (4 \pi)^2  u^2 \over 9}  
\left|  {\alpha_s \over t} + \alpha_s \tan^2 \phi P_t^{F'} 
\right|^2 \\
|{\cal M}(u_L\bar{s_R}\rightarrow u_L\bar{s_R})|^2 &  
= & |{\cal M}(u_R\bar{s_L}\rightarrow u_R\bar{s_L})|^2 =
{2 (4 \pi)^2 s^2 \over 9} \left| {\alpha_s \over t} - \alpha_s P_t^{F'} 
\right|^2 
\end{eqnarray}




\newpage

\begin{thebibliography}{99}


\bibitem{ETC}
      S. Dimopolous and L. Susskind, {\em Nucl. Phys.} {\bf B155} 237 (1979);
      E. Eitchen and K. Lane, {\em Phys. Lett.} {\bf B90} 125 (1980).


\bibitem{TC} S. Weinberg, {\em Phys. Rev.} {\bf D19} 1277 (1979);
      L. Suskind, {\em Phys. Rev.} {\bf D20} 2619 (1979);
      E. Farhi and L. Susskind, {\em Phys. Report} 74 No.3 277 (1981).

\bibitem{tc}  Y. Nambu, ``New Theories In Physics", Proc. XI Warsaw
Symposium on Elementary Particle Physics, (ed. Z. Adjuk {\it et al.}, publ.
World  Scientific, Singapore, 1989);
V.A. Miransky, M.Tanabashi and M. Yamawaki, {\em Phys. Lett.} {\bf
B221} (1989) 177; R.R. Mendel and V.A. Miransky, 
{\em Phys. Lett.} {\bf B 268} 384 (1991);
W.A. Bardeen, C.T. Hill and M.Lindner, {\em Phys. Rev.} {\bf D41}
1647 (1990).

\bibitem{tcol}C. T. Hill, {\em Phys. Lett.~}{\bf B345}, 483 (1995). 

\bibitem{king} S. F. King, {\em Phys. Rev.~}{\bf D45}, 990 (1992).

\bibitem{tseesaw} B. Dobrescu and C. T. Hill,{\em ~Phys. Rev. Lett.} 
         {\bf 81} 2634 (1998); 
         R.S. Chivukula, B.A. Dobrescu, H. Georgi and C.T. Hill, 
         {\bf hep-ph/9809470}. 

\bibitem{fuseesaw}  G. Burdman and N. Evans;
{\em Phys. Rev.} {\bf D59} (1999) 115005.

\bibitem{ctsm} R. S. Chivukula and H. Georgi, {\em Phys. Lett.} {\bf 188B}
99 (1987).

\bibitem{georgi} H. Georgi, ``Technicolor and Families", Proc. 1990
International
         Workshop On Strong Coupling Gauge Theories And Beyond, (ed. T. Muta
         and K. Yamawaki, publ. World Scientific, Singapore,1991);
          H. Georgi, {\em Nucl. Phys.} {\bf B416} 699
         (1994).


\bibitem{mixing} R.S. Chivukula and J. Terning; 
{\em Phys. Lett.} {\bf B385} (1996) 209.

\bibitem{zbounds} G. Burdman, R. S. Chivukula and N. Evans, 
{\em Phys. Rev.~}{\bf D61} 035009 (2000).

\bibitem{unicol} R.S. Chivukula, A.G. Cohen and  E.H. Simmons 1996. 8pp. 
{\em Phys. Lett.} {\bf B380} (1996) 92. 

\bibitem{NJL} Y. Nambu and  G. Jona-Lasinio;
{\em Phys. Rev.} {\bf 122}  (1961) 345.

\bibitem{simmons} E. H. Simmons, {\em Phys. Rev.} {\bf D55} 1678 
(1997).

\bibitem{bertram} I. Bertram and E. H. Simmons, {\em Phys. Lett.} {\bf
    B443} 347 (1998).

\bibitem{d0kdata} B. Abbot {\em et al.}, the D0 collaboration, 
{\em Phys. Rev. Lett.~}{\bf 89} 2457 (1999).

\bibitem{cdfst} P. Koehn, the CDF Collaboration, FERMILAB-CONF-99/306-E. 
Published Proceedings International Europhysics Conference on
High-Energy Physics (EPS-HEP 99), Tampere, Finland, July 15-21, 1999. 

\bibitem{scott} T. Stelzer, Z. Sullivan and S. Willenbrock, {\em Phys. 
Rev.~}{\bf D58} 094021 (1998).

\bibitem{randall} L. Randall, {\em Nucl. Phys.} {\bf B403} 122 (1993).




\bibitem{apvexp} S. C. Bennet and C. E. Wieman, {\em Phys. Rev. Lett.}, 
{\bf 82} 2484 (1999). 

\bibitem{apvth} C. Caso {\em et al.}, {\em Eur.~Phys.~J.~}{\bf C3} 1 (1998). 
The most recent value of $Q_W(Cs)$ can be found in 1999 update of the PDG at
http://pdg.lbl.gov. 



\end{thebibliography}
\end{document}